\documentclass[preprint,12pt,3p]{elsarticle}

\usepackage{amssymb}
\usepackage{graphicx}
\usepackage{amsfonts}
\usepackage{subfigure}
\usepackage{amsthm}

\journal{Phys Letters B}

\begin{document}

\begin{frontmatter}

\title{Non-commutative and commutative vacua effects in a scalar torsion scenario}

\author[label1]{Haidar Sheikhahmadi\corref{cor1}}
\address[label1]{Department of Physics, Faculty of Science, University of Kurdistan,  Sanandaj, Iran.}
\cortext[cor1]{I am corresponding author}
\ead{h.sh.ahmadi@gmail.com}

\author[label2]{Ali Aghamohammadi}
\address[label2]{Sanandaj Branch, Islamic Azad University, Sanandaj, Iran.}
\ead{a.aghamohamadi@iausdj.ac.ir }

\author[label1]{Khaled Saaidi}
\ead{ksaaidi@uok.ac.ir}
\begin{abstract}
In this work, the effects of non-commutative and commutative vacua on the phase space generated by a scalar field in a scalar torsion scenario are investigated. For both classical and quantum regimes, the commutative and non-commutative cases are compared. To take account the effects of non-commutativity, two well known non-commutative parameters, $\theta$ and $\beta,$ are introduced. It should be emphasized, the effects of $\beta$ which is related to momentum sector has more key role in comparison to $\theta$ which is related to space sector. Also the different boundary conditions and mathematical interpretations of non-commutativity are explored.
\end{abstract}


\end{frontmatter}


\section{Introduction}
\label{sec1}
\indent
In this work our aim is studying a non-commutative model of scalar torsion gravity.\\
Recently some astrophysical observations have shown that the Universe undergoing an accelerated phase era. To justify this unexpected result, scientists have proposed some different models such as, scalar field models~\citep{SF, SF1, SF2, Sf3} and modify theories of  gravity ~\cite{MDG, MDG1, MDG2, MDG3}. For the latter proposal one can deal with telleparallel equivalent of general relativity~\cite{R1, TPE, TPE1, TPE2}, in which the field equations are second order~\cite{O2}. In addition in this scenario the Levi-Civita connections replaced by Weitzenb\"{o}ck connection, where has no curvature but only torsion~\cite{OT}.\\
It is obvious that for the first time, the non-commutative formalism between the space-time coordinate was introduced by~\cite{FNC}. Also the geometrical concept based on this model recently attracts more interesting namely non-commutative geometry~\cite{GNC, GNC1, GNC2, GNC3}. It is notable the recent investigations of string theory, supersymmetry, M-theory and so on~\cite{MT, MT1}, motivated scientists to study classical and quantum cosmology in such frame. The effects of non-commutativity in cosmology have been investigated by two well-known models, i.e. minisuperspace~\cite{MSS, MSS1} and phase space~\cite{PHS}, while the geometrical structure of the underlying space-time unchanged~\cite{MALEK}.
In this work our means is that to build up a non-commutative scenario by means of a deformation achieved by Moyal product~\cite{MP}, for a scalar torsion gravity~\cite{RM} in both classical and quantum levels. Although the  non-minimal coupling term has a richer structure and experiencing the phantom-divide crossing  and so on \cite{R1, R2, R3}, but we want to consider the simplest form, minimal quintessence-like, of a scalar torsion scenario in comparison to standard quintessence scenario for this investigation.\\
The organization of this work is as follows: In Sec.\ref{sec2}, a brief review about scalar $f(T)$ gravity cosmology and general properties of the model are discussed. In Sec.\ref{sec3}, the results of our investigations for scalar torsion gravity are discussed in classical level for both commutative and non-commutative frames. The Sec.\ref{sec4}, is devoted to the same details of Sec.\ref{sec3} but in quantum level. And at last the Sec.\ref{sec5}, is concerned with the conclusion and discussion.
\section{General framework}
\label{sec2}
\indent
 The telleparallel theory of gravity is defined in the Weitzenb\"{o}ck's space-time by the following line element
\begin{equation}\label{el}
dS^{2}=N^2dt^2-a^2(t)\delta_{ij}dx^idx^j\; ,
\end{equation}
where $N$ is the lapse function. Also it is considerable that, theory can be described in the tangent space, which allows us to rewrite the line element (\ref{el}) as
\begin{eqnarray}
dS^{2} &=&g_{\mu\nu}dx^{\mu}dx^{\nu}=\eta_{ij}\theta^{i}\theta^{j}\label{1}\; ,\\
dx^{\mu}& =&e_{i}^{\;\;\mu}\theta^{i}\; , \; \theta^{i}=e^{i}_{\;\;\mu}dx^{\mu}\label{2}\; ,
\end{eqnarray}
where $\eta_{ij}=diag[1,-1,-1,-1]$ and $e_{i}^{\;\;\mu}e^{i}_{\;\;\nu}=\delta^{\mu}_{\nu}$ or  $e_{i}^{\;\;\mu}e^{j}_{\;\;\mu}=\delta^{j}_{i}$. and the matrix $e^{a}_{\;\;\mu}$ are called tetrads that indicate the dynamic fields of the theory.
\par
According to theses fields, the Weitzenb\"{o}ck's connection is defined as
\begin{eqnarray}
\Gamma^{\alpha}_{\mu\nu}=e_{i}^{\;\;\alpha}\partial_{\nu}e^{i}_{\;\;\mu}=-e^{i}_{\;\;\mu}\partial_{\nu}e_{i}^{\;\;\alpha}\label{co}\; ,
\end{eqnarray}
that to be used for construction the main geometrical objects of the space-time.
The components of the tensor torsion and the contorsion are defined respectively as
\begin{eqnarray}
T^\rho_{\verb| |\mu\nu}\equiv e^{~\rho}_l
\left( \partial_\mu e^{l}_{~\nu} - \partial_\nu e^{l}_{~\mu} \right)\,,
\label{eq:2.2} \\
K^{\mu\nu}_{\verb|  |\rho}\equiv
-\frac{1}{2}
\left(T^{\mu\nu}_{\verb|  |\rho} - T^{\nu \mu}_{\verb|  |\rho} -
T_\rho^{\verb| |\mu\nu}\right)\,.
\label{eq:2.3}
\end{eqnarray}
It was defined  a new tensor $S_\rho^{\verb| |\mu\nu}$, to obtain  the scalar equivalent to the curvature scalar of general relativity i.e. Ricci scalar, that is as
\begin{equation}
S_\rho^{\verb| |\mu\nu}\equiv \frac{1}{2}
\left(K^{\mu\nu}_{\verb|  |\rho}+\delta^\mu_\rho
T^{\alpha \nu}_{\verb|  |\alpha}-\delta^\nu_\rho
T^{\alpha \mu}_{\verb|  |\alpha}\right).
\label{eq:2.5}
\end{equation}
Hence, the torsion scalar is defined by  the following contraction
\begin{equation}
T \equiv S_\rho^{\verb| |\mu\nu} T^\rho_{\verb| |\mu\nu}.
\label{eq:2.4}
\end{equation}
In studying the scalar torsion model instead of non-minimal coupling scenario~\cite{R1, R3}, the minimal coupling action of the theory is defined by generalizing  the teleparallel theory, as~\cite{RM}
\begin{equation}\label{A1}
\mathcal{A}=\int
d^4x |e|\Bigl[{\xi {T}}-\zeta\frac{1}{2}\eta^{ij}e_{i}^{~\mu}e_{j}^{~\nu}{\nabla_{\mu}}\phi{\nabla_{\nu}}\phi-V(\phi)\Bigr],
\end{equation}
where $|e|=\sqrt(-g)$ and $T$ is the torsion scalar, $\xi$ and $\zeta$ are constant.
Let us choose the following set of diagonal tetrads related to the metric (\ref{el}) as
\begin{eqnarray}
\left[e^{a}_{\;\;\mu}\right]=diag\left[N,a,a,a\right]\;, \label{matrixtype3}
\end{eqnarray}
the determinant of the matrix (\ref{matrixtype3}) is $e=Na^3$. The components of the torsion tensor (\ref{eq:2.2}) for the tetrads (\ref{matrixtype3}) are given by
\begin{eqnarray}
T^{1}_{\;\;01}=\frac{\dot{a}}{Na}=T^{2}_{\;\;02}=T^{3}_{\;\;03}\;,\label{torsiontype3}
\end{eqnarray}
and the components of the corresponding contorsion are
\begin{eqnarray}
K^{01}_{\;\;\;\;1}=\frac{\dot{a}}{Na}=K^{02}_{\;\;\;\;2}=K^{03}_{\;\;\;\;3}\;.\label{contorsiontype3}
\end{eqnarray}
The components of the tensor $S_{\alpha}^{\;\;\mu\nu}$, in (\ref{eq:2.5}),  are given by
\begin{eqnarray}
S_{1}^{\;\;10}=(\frac{\dot{a}}{Na})=S_{2}^{\;\;20}=\,S_{3}^{\;\;30}\;.\label{tensortype3}
\end{eqnarray}
By using the components (\ref{torsiontype3}) and (\ref{tensortype3}),  the torsion scalar (\ref{eq:2.4}) is given by $$T=-6\frac{\dot{a}^{2}}{(Na)^{2}}\,.$$
Substituting Eq.(\ref{matrixtype3}) into the action(\ref{A1}) the Lagrangian density can be achieved as follows
\begin{equation}\label{A3}
\mathcal{L}={{N}}{{a}^3}\Bigl(-{6}{\xi}\frac{\dot{a}^{2}}{(Na)^2}+\frac{\zeta}{2N^2}\dot{\phi}^{2}-V(\phi)\Bigr),
\end{equation}
For more convenience the above constants $\xi$ and $\zeta$ can be considered as $\xi=1/6,~\zeta=1/2$.
Using a new set of variables,
\begin{equation}\label{variables}
x=\frac{a^{2}}{2}\cosh\phi,~y=\frac{a^{2}}{2}\sinh\phi\,,
\end{equation}
where $a^{2}=2(x-y)e^{\phi}\, ,$
one can rewrite the above Lagrangian density as follows
\begin{equation}\label{A3}
\mathcal{L}=(\dot{y}^{2}-\dot{x}^{2})-4(x-y)e^{\phi}V(\phi)\,.
\end{equation}
Thence, the corresponding Hamiltonian density is
\begin{equation}\label{A3}
\mathcal{H}\equiv\sum\limits_{{\alpha}}\dot{x}^{\alpha}\frac{\partial\mathcal{L}}{\partial \dot{x}^{\alpha}}-\mathcal{L}=\frac{1}{2}(\frac{1}{2}P_{y}^{2}-\frac{1}{2}P_{x}^{2})+4(x-y)e^{\phi}V(\phi)\,.
\end{equation}
where $V(\phi)=2V_0\exp{[-\phi]}$ and $V_0$ is a constant.
\section{The cosmological evolution in classical regime}
\label{sec3}
\indent
It is clear the classical solutions of a specific Hamiltonian can be easily yielded. However we want to  inspect the effects of non-commutativity in classical level, then  compare our results with commutative case.
\subsection{Commutative algebra}
\label{subsec1}
\indent
It is well known the Poisson brackets between components of the classical phase space variables are as
\begin{equation}\label{J}
\left\{x_i,x_j\right\}=\left\{p_i,p_j\right\}=0,\hspace{.5cm}\left\{x_i,p_j\right\}=\delta_{ij},
\end{equation}
where $x_i(i=1,2)=x,y$ and $p_i(i=1,2)=p_x, p_y.$ Assuming
${N}={1}/{a}$, the equations of motion to be as
\begin{equation}\label{K}
\dot{x}=\left\{x,{\cal
H}\right\}=-\frac{p_x}{2},\hspace{.5cm}\dot{p_x}=\left\{p_x,{\cal
H}\right\}=-8V_0\, ,
\end{equation}
\begin{equation}\label{L}
\dot{y}=\left\{y,{\cal
H}\right\}=\frac{p_y}{2},\hspace{.5cm}\dot{p_y}=\left\{p_y,{\cal
H}\right\}=8V_0\, .
\end{equation}
Integrating the above equations, get
\begin{equation}\label{M}
x(t)=4V_0t^2-p_{0x}t+x_0,\hspace{.5cm}p_x(t)=-8V_0t+p_{0x}\,
\end{equation}
\begin{equation}\label{N}
y(t)=4V_0t^2+p_{0y}t+y_0,\hspace{.5cm}p_y(t)=8V_0t+p_{0y}\, ,
\end{equation}
Where $x_0,\, y_0,\,p_{0x}$ and $p_{0y}$ are integration constants. In addition the constraint equation between them, by using the zero energy condition, ${\cal{H}}\equiv 0$, yields
\begin{equation}\label{O}
p_{0x}^2-p_{0y}^2=-16V_0 (y_0-x_0)\, .
\end{equation}
It is clear the Eqs.(\ref{M}) and (\ref{N}) have the same  form of the equation motion of a particle with a constant acceleration. one can apply the condition $x>0,$ with the bound $p_{0x}^2-16V_0x_0<0$ to obtain the constraint $p_{0y}^2-16V_0y_0<0$ from relation (\ref{O}), which indicates that $y>0.$ So only half of minisuperspace $x>y>0$ is covered by dynamical variables. The evolution of scale factor and scalar field by combination Eqs.(\ref{variables}), (\ref{M}) and (\ref{N}) are given as follows
\begin{equation}\label{O1}
{a}(t)=\left(8|p_{0x}|(8V_0t^3+2x_0t)\right)^{1/4}~,
\end{equation}
\begin{equation}\label{O2}
\phi(t)=\frac{1}{2}\ln\left(\frac{8V_0t^2+2x_0}{2|p_{0x}|t}\right)\,,
\end{equation}
where we suppose $x_0=y_0$ and $p_{0x}=p_{0y}$, in agreement with Eq.(\ref{O}). Based on Eq.(\ref{O1}), $\ddot{a}<0$ so the Universe is in a decelerated phase epoch. According to the Eq.(\ref{variables}), one can define an effective scale factor, $a^{2}_{eff}=a^2e^{-\phi}$, which is equal to
\begin{equation}\label{effective-class}
a^{2}_{eff}=2(x-y)=4|p_{0x}|t\, .
\end{equation}
Hence, it is obvious that the above equation indicates the radiation dominated era.
\subsection{Non-commutative algebra}
\label{subsec2}
\indent
This subsection is concerned with the effects of non-commutativity in a classical cosmology.
To investigate the influences of non-commutativity in classical level, one requires the star product law, the Poisson and the Moyal brackets which were discussed in detail at~\cite{MALEK, MP}. The Moyal
product law between two arbitrary functions of phase space variables, namely
$\mathfrak{F}^a=(x^i,p^j)$ for $i=1, \cdots, l$ and $j=l+1, \cdots, 2l$,
are defined as~\cite{MP}
\begin{eqnarray}\label{A1}
(f \ast g)(\mathfrak{F})=\exp\left[\frac{1}{2}\alpha^{ab}\partial_a^{(1)}\partial_b^{(2)}\right]f(\mathfrak{F}_1)g(\mathfrak{F}_2)
{\biggr|}_{\mathfrak{F}_1=\mathfrak{F}_2=\mathfrak{F}}\, ,
\end{eqnarray}
so that
\begin{equation}\label{A2}
(\alpha_{ab})=\left(%
\begin{array}{cc}
\theta_{ij} & \delta_{ij}+\sigma_{ij} \\-\delta_{ij}-\sigma_{ij}& \beta_{ij} \\
\end{array}
\right)\, ,
\end{equation}
where $a, b=1, 2, \cdots, 2l$, $\theta_{ij}$ and $\beta_{ij}$ are
the elements of real and antisymmetric matrices,
$\sigma_{ij}$ is a symmetric matrix and dimension of
the classical phase space is $2l$. Thence the deformed
Poisson brackets are defined as
\begin{equation}\label{A3}
\{f,g\}_\alpha=f\ast g-g\ast f\, .
\end{equation}
It is well known, the Poisson brackets between the phase space coordinate could be written as
\begin{equation}\label{T}
\{x_i,x_j\}_\alpha=\theta_{ij},\hspace{.5cm}\{x_i,p_j\}_\alpha=\delta_{ij}+\sigma_{ij},\hspace{.5cm}\{p_i,p_j\}_\alpha=\beta_{ij}\, .
\end{equation}
To obtain the usual poisson bracket forms (\ref{J})
\begin{equation}\label{V}
\{x'_i,x'_j\}=\theta_{ij},\hspace{.5cm}\{x'_i,p'_j\}=\delta_{ij}+\sigma_{ij},\hspace{.5cm}\{p'_i,p'_j\}=\beta_{ij}~,
\end{equation}
one can make a transformation as~\cite{17M}
\begin{equation}\label{U}
x'_i=x_i-\frac{1}{2}\theta_{ij}p^j,\hspace{.5cm}p'_i=p_i+\frac{1}{2}\beta_{ij}x^j,
\end{equation}
where $\sigma_{ij}=-\frac{1}{8}\left(\theta_i^k\beta_{kj}+\beta_i^k\theta_{kj}\right)$.
By considering $\theta_{12}=\theta$ and $\beta_{12}=\beta$, one able to show that only following Poisson brackets could be exist
\begin{equation}\label{W}
\left\{x',y'\right\}=\theta,\hspace{.5cm}\left\{x',p'_x\right\}=\left\{y',p'_y\right\}=1-\theta
\beta/4,\hspace{.5cm}\left\{p'_x,p'_y\right\}=\beta\, .
\end{equation}
In non-commutative case, the Hamiltonian takes a similar form as classical ones,
\begin{equation}\label{X11}
{\cal H}_{nc}=\frac{1}{2}\Bigl[-\frac{1}{2}p'^2_x+\frac{1}{2}p'^2_y\Bigr]+8V_0\left(x'-y'\right),
\end{equation}
but it should be noted in this case, the dynamical variables satisfy the deformed Poisson brackets (\ref{U}), therefore Eq.(\ref{X11}) is reduced to
\begin{equation}\label{X}
{\cal H}_{nc}=\frac{1}{2}\Bigl[\frac{p^2_y-p^2_x}{2}+\frac{\beta^2}{8}(x^2-y^2)-\frac{\beta}{2}(xp_y+yp_x)-4{V_0\theta}(p_x+p_y)
\Bigr]+8V_0\left(x-y\right)~.
\end{equation}
Hence, the equations of motion are achieved as
\begin{eqnarray}\label{Y}
2\dot{x}&=&\left\{x,{\cal H}_{nc}\right\}=-p_x-\frac{\beta}{2}y-4\theta V_0
,\hspace{.25cm}2\dot{p_x}=\left\{p_x,{\cal H}_{nc}\right\}=-\frac{\beta^2}{4}x+\frac{\beta}{2}p_y-16V_0~, \cr
2\dot{y}&=&\left\{y,{\cal H}_{nc}\right\}=p_y-\frac{\beta}{2}x-4{\theta V_0},\hspace{.25cm}2\dot{p_y}=\left\{p_y,{\cal
H}_{nc}\right\}=\frac{\beta^2}{4}y+\frac{\beta}{2}p_x+16V_0\, .
\end{eqnarray}
Integrating above equations, and after some algebra the dynamical variables are attained as
\begin{eqnarray}\label{AB}
x(t)&=&Ae^{\beta t}+Be^{-\beta t}+Ct+D_1\, ,\cr
y(t)&=&-Ae^{\beta t}+Be^{-\beta t}+Ct+D_2\, ,
\end{eqnarray}
where $C\equiv8(1-\theta \beta/4)V_0/\beta$ and $A$, $B$, $D_1$ and
$D_2$ are integration constants which their values are restricted to
satisfy the constraint equation ${\cal H}_{nc}=0$, that is
\begin{equation}\label{ABC}
\beta^2AB=4C(D_1-D_2)\, .
\end{equation}
Now with a calculation such as the preceding case concern with the scale factor, one can obtain
\begin{equation}\label{scalenc}
a^{2}_{eff}(t)=2\left[2Ae^{\beta t}+\frac{\beta^3AB}{32V_0(1-\frac{\theta\beta}{4})}\right]\, .
\end{equation}
To obtain coefficient $B$, if we impose the condition $a_{eff}(0)=0$, the Eq.(\ref{scalenc}) reduces to
\begin{equation}\label{redscalenc}
a^{2}_{eff}(t)=4A\big[e^{\beta t}-1\big]\, .
\end{equation}
 The asymptotically behaviour interpretation about above equation is as follows
\begin{itemize}
\item For the early time, expanding the exponential function it is clear $a_{eff}(t)\propto\sqrt{t}$, which is in agreement with radiation dominated epoch. By the way this case is similar to commutative ones.
\item At the late time, the effective scale factor is proportional to $e^{\beta t/2}$, which behaves such as accelerated de Sitter Universe, thence it is expected that coefficient $\beta$ plays the role of $\Lambda$ cosmological constant.
\end{itemize}
Therefore the importance of the existence of scalar field, $\phi$, in a non-commutative scalar torsion gravity, is that the effective scale factor can justify the accelerated Universe.
\section{The cosmological evolution in quantum regime}
\label{sec4}
\indent
This section is concern with the quantization of the cosmological model given by action (\ref{A1})
for the free potential case in which the canonical quantization gives the Wheeler-De Witt (WD) equation, ${\cal H} \Psi=0$. For more explanations we refer the reader to~\cite{WD}.
\subsection{Commutative algebra}
\label{subsec3}
\indent
It is well known, by means of the  operator forms as $ p_a\rightarrow -i\partial_a$ and
$p_\chi\rightarrow -i\partial_\chi$ the Hamiltonian (\ref{X}) can act as an operator. Considering a particular factor ordering, the corresponding WD equation is
\begin{equation}\label{D1}
\left[\frac{\partial^2}{\partial x^2}-\frac{\partial^2}{\partial
y^2}\right]\Psi(x,y)=0\, .
\end{equation}
Assuming the following change of variables
\begin{equation}\label{D2}
x=\rho\cosh\phi\hspace{.75cm}{\rm and} \hspace{.75cm}
y=\rho\sinh\phi\hspace{.1cm}\, ,
\end{equation}
the differential equation (\ref{D1}) can be rewritten as
\begin{equation}\label{D3}
\left(\frac{\partial^2}{\partial
\rho^2}+\frac{1}{\rho}\frac{\partial}{\partial
\rho}-\frac{1}{\rho^2}\frac{\partial^2}{\partial
\phi^2}\right)\Psi(\rho,\phi)=0\, ,
\end{equation}
If one consider the following product solution for above equation
\begin{equation}\label{D4}
\Psi(\rho,\phi)=\psi(\rho)e^{2i\tilde{\alpha}\phi}\, ,
\end{equation}
where $\alpha$ is a constant, the Eq.(\ref{D3}) is obtained as
\begin{equation}\label{D5}
\frac{d^2\psi}{d\rho^2}+\frac{1}{\rho}\frac{d\psi}{d\rho}+4\frac{\tilde{\alpha}^2}{\rho^2}\psi=0\, .
\end{equation}
By considering suitable boundary condition, the eigenfunction of above equation could be written as
\begin{equation}\label{D55}
\psi(\rho)={\cal{R}}\cos(2 \tilde{\alpha} \ln \rho)\, ,
\end{equation}
where ${\cal{R}}$ is integration constant. Therefore, by using the offered solution (\ref{D4}) the wave packet corresponding to Eq.(\ref{D55}) is as
\begin{equation}\label{D72}
\Psi(\rho,
\phi)=\int^{+\infty}_{-\infty}w_{\tilde{\alpha}}\psi_{\tilde{\alpha}}(\rho)e^{2i\tilde{\alpha}\phi}d\tilde{\alpha}\,
,
\end{equation}
where $w_{\tilde{\alpha}}$ can be introduce as the shifted Gaussian
weight function with constants $b$ and $c$ ~\cite{9M}.

\subsection{Non-commutative algebra}
\label{subsec4}
\indent
The noncommutative WD equation corresponding to relation (\ref{X}), for $V=0$, is as
\begin{equation}\label{D9}
\left[(\partial^2_x-\partial^2_y)+
i\beta(y\partial_x+x\partial_y)+\frac{\beta^2}{4}(x^2-y^2)\right]\Psi(x,y)=0,
\end{equation}
that, with the change of variables (\ref{D2}), reads
\begin{equation}\label{D10}
\left[\Big(\frac{\partial^2}{\partial
\rho^2}+\frac{1}{\rho}\frac{\partial}{\partial
\rho}-\frac{1}{\rho^2}\frac{\partial^2}{\partial \phi^2}\Big)+
4i\beta\frac{\partial}{\partial
\phi}+4\beta^2\rho^2\right]\Psi=0\, .
\end{equation}
By using the offered product solution (\ref{D4}), the Eq.(\ref{D10}) reduces to
\begin{equation}\label{D11}
\frac{d^2\psi}{d\rho^2}+\frac{1}{\rho}\frac{d\psi}{d\rho}+4\left(\frac{{\tilde{\alpha}}^2}{\rho^2}+\frac{\beta^2}{16}\rho^2-\frac{1}{2}\beta{\tilde{\alpha}}\right)\psi=0\,
,
\end{equation}
Before any discussion about the solution of Eq.(\ref{D11}), let's enumerate some features of it. As the first note, one can see by placing $\beta=0$ it reduces to commutative case. Secondly it is interesting to note that when the chosen background is flat FLRW, the effect of $\theta$ does not appear. So the $\beta$ coefficient, plays the more key role rather than $\theta$. Solving Eq.(\ref{D11})leads to a solution based on a combination of hypergeometry
and associated Laguerre functions, which by reparametrization can be rearranged as Whittaker functions, $M_{\mu~,\nu}$ and $W_{\mu~,\nu}$, as
\begin{equation}\label{D12}
\psi_{\tilde{\alpha}}\left(\rho\right)=\rho^{-1}\left[A_{\tilde{\alpha}}M_{\mu\,
,\nu}\left(2 i\beta \rho^{2}\right) +B_{\tilde{\alpha}}W_{\mu\,
,\nu}\left( 2 i\beta \rho^{2}\right)\right]\, .
\end{equation}
Where $A_{\tilde{\alpha}}$ and $B_{\tilde{\alpha}}$ are superposition coefficients, $\mu=i{\tilde{\alpha}}/4$ and $\nu=i{\tilde{\alpha}}$. It should be noted where the argument of both Whittaker functions is imaginary, therefore even in classical forbidden areas they are convergent. But in a special case which $\beta$ gets the imaginary values the Whittaker functions are divergent. In addition in this case the Whittaker functions are quickly damped as $\rho$ increases. Using Eqs. (\ref{D4}) and (\ref{D12}), the solution of differential equation (\ref{D11}), one gets
\begin{equation}\label{D13}
\Psi(\rho,\phi)=\rho^{-1}\int^\infty_{-\infty}e^{-b({\tilde{\alpha}}-c)^2}
M_{\frac{i{\tilde{\alpha}}}{4}~,i{\tilde{\alpha}}}\left(2i\beta\rho^{2}\right)e^{2i{\tilde{\alpha}}\phi}d{\tilde{\alpha}}~.
\end{equation}
\section{Conclusion and Discussion}
\label{sec5}
The effects of non-commutative and commutative vacua on the phase space generated by a scalar field in a scalar torsion scenario have been investigated. For both classical and quantum regimes, the commutative and non-commutative cases have been compared. The asymptotically behaviour interpretation as to effective scale factor has shown that, for the early time, $a_{eff}(t)\propto\sqrt{t}$, which is in agreement with radiation dominated epoch. It is notable for such era the non-commutative case is similar with commutative ones. Also At the late time, the effective scale factor is proportional to $e^{\beta t/2}$, which behaves as accelerated de Sitter Universe, thence it was expected that coefficient $\beta$ plays the role of $\Lambda$ cosmological constant.
It is also notable, in a non-commutative scalar torsion gravity because of the existence of scalar field, $\phi$, the effective scale factor justifies the accelerated Universe. It was understood our results for commutative and non-commutative cases in scalar torsion cosmology are in good agreement in comparison to $f(R)$ gravity~\cite{MALEK}.
\section{Acknowledgment}
We would like to thank Behrooz Malekolkalami for useful comments and a clarifying discussion about non-commutativity.


\bibliographystyle{elsarticle-num}

\bibliography{samplerevised}

\begin{thebibliography}{10}
\expandafter\ifx\csname url\endcsname\relax
  \def\url#1{\texttt{#1}}\fi
\expandafter\ifx\csname urlprefix\endcsname\relax\def\urlprefix{URL }\fi
\expandafter\ifx\csname href\endcsname\relax
  \def\href#1#2{#2} \def\path#1{#1}\fi

\bibitem{SF}
C.~Brans, R.~H. Dicke, Mach's principle and a relativistic theory of
  gravitation, Phys. Rev. 124~(0) (1961) 925--935.
\newblock \href {http://dx.doi.org/http://dx.doi.org/10.1103/PhysRev.124.925}
  {\path{doi:http://dx.doi.org/10.1103/PhysRev.124.925}}.

\bibitem{SF1}
J.~Khoury, A.~Weltman, {Chameleon cosmology}, Annalen der Physik 69~(15) (2004)
  044026.
\newblock \href
  {http://dx.doi.org/http://link.aps.org/doi/10.1103/PhysRevD.69.044026}
  {\path{doi:http://link.aps.org/doi/10.1103/PhysRevD.69.044026}}.

\bibitem{SF2}
D.~F. Mota, J.~D. Barrow, Varying alpha in a more realistic universe, Physics
  Letters B 581~(3–4) (2004) 141--146.
\newblock \href
  {http://dx.doi.org/http://dx.doi.org/10.1016/j.physletb.2003.12.016}
  {\path{doi:http://dx.doi.org/10.1016/j.physletb.2003.12.016}}.

\bibitem{Sf3}
K.~Saaidi, A.~Mohammadi, H.~Sheikhahmadi, $\ensuremath{\gamma}$ parameter and
  solar system constraint in chameleon-brans-dicke theory, Phys. Rev. D 83~(6)
  (2011) 104019.
\newblock \href
  {http://dx.doi.org/http://dx.doi.org/10.1103/PhysRevD.83.104019}
  {\path{doi:http://dx.doi.org/10.1103/PhysRevD.83.104019}}.

\bibitem{MDG}
D.~Wands, Extended gravity theories and the einstein--hilbert action, Classical
  and Quantum Gravity 11~(1) (1994) 269.
\newblock \href {http://dx.doi.org/10.1088/0264-9381/11/1/025}
  {\path{doi:10.1088/0264-9381/11/1/025}}.

\bibitem{MDG1}
S.~Nojiri, S.~D. Odintsov, Modified $f(r)$ gravity consistent with realistic
  cosmology: From a matter dominated epoch to a dark energy universe, Phys.
  Rev. D 74 (2006) 086005.
\newblock \href
  {http://dx.doi.org/http://dx.doi.org/10.1103/PhysRevD.74.086005}
  {\path{doi:http://dx.doi.org/10.1103/PhysRevD.74.086005}}.

\bibitem{MDG2}
A.~Guarnizo, L.~Castañeda, J.~Tejeiro, Boundary term in metric f (r) gravity:
  field equations in the metric formalism, General Relativity and Gravitation
  42~(11) (2010) 2713--2728.
\newblock \href {http://dx.doi.org/10.1007/s10714-010-1012-6}
  {\path{doi:10.1007/s10714-010-1012-6}}.

\bibitem{MDG3}
K.~Saaidi, A.~Aghamohammadi, B.~Sabet, O.~Farooq, Ghost dark energy in f(r)
  model of gravity, International Journal of Modern Physics D 21~(06) (2012)
  1250057.
\newblock \href {http://dx.doi.org/10.1142/S0218271812500575}
  {\path{doi:10.1142/S0218271812500575}}.

\bibitem{R1}
C.-Q. Geng, C.-C. Lee, E.~N. Saridakis, Y.-P. Wu, “teleparallel” dark
  energy, Physics Letters B 704~(5) (2011) 384 -- 387.
\newblock \href
  {http://dx.doi.org/http://dx.doi.org/10.1016/j.physletb.2011.09.082}
  {\path{doi:http://dx.doi.org/10.1016/j.physletb.2011.09.082}}.

\bibitem{TPE}
E.~V. Linder, Einstein's other gravity and the acceleration of the universe,
  Phys. Rev. D 81 (2010) 127301.
\newblock \href
  {http://dx.doi.org/http://dx.doi.org/10.1103/PhysRevD.81.127301}
  {\path{doi:http://dx.doi.org/10.1103/PhysRevD.81.127301}}.

\bibitem{TPE1}
K.~Bamba, C.-Q. Geng, C.-C. Lee, L.-W. Luo, Equation of state for dark energy
  in f (t) gravity, Journal of Cosmology and Astroparticle Physics 2011~(01)
  (2011) 021.
\newblock \href {http://dx.doi.org/:doi:10.1088/1475-7516/2011/01/021}
  {\path{doi::doi:10.1088/1475-7516/2011/01/021}}.

\bibitem{TPE2}
A.~Aghamohammadi, Holographic f(t) gravity model, Astrophysics and Space
  Science 352~(1) (2014) 1--5.
\newblock \href {http://dx.doi.org/10.1007/s10509-014-1912-0}
  {\path{doi:10.1007/s10509-014-1912-0}}.

\bibitem{O2}
G.~R. Bengochea, R.~Ferraro, Dark torsion as the cosmic speed-up, Phys. Rev. D
  79 (2009) 124019.
\newblock \href
  {http://dx.doi.org/http://dx.doi.org/10.1103/PhysRevD.79.124019}
  {\path{doi:http://dx.doi.org/10.1103/PhysRevD.79.124019}}.

\bibitem{OT}
R.~Ferraro, F.~Fiorini, Modified teleparallel gravity: Inflation without an
  inflaton, Phys. Rev. D 75~(5) (2007) 084031.
\newblock \href
  {http://dx.doi.org/http://dx.doi.org/10.1103/PhysRevD.75.084031}
  {\path{doi:http://dx.doi.org/10.1103/PhysRevD.75.084031}}.

\bibitem{FNC}
H.~S. Snyder, Quantized space-time, Phys. Rev. 71~(0) (1947) 38--41.
\newblock \href {http://dx.doi.org/http://dx.doi.org/10.1103/PhysRev.71.38}
  {\path{doi:http://dx.doi.org/10.1103/PhysRev.71.38}}.

\bibitem{GNC}
A.~Connes, A short survey of noncommutative geometry, Journal of Mathematical
  Physics 41~(6) (2000) 3832--3866.
\newblock \href {http://dx.doi.org/http://dx.doi.org/10.1063/1.533329}
  {\path{doi:http://dx.doi.org/10.1063/1.533329}}.

\bibitem{GNC1}
A.~Connes, An analogue of the thom isomorphism for crossed products of a c∗
  algebra by an action of r, Advances in Mathematics 39~(1) (1981) 31 -- 55.
\newblock \href
  {http://dx.doi.org/http://dx.doi.org/10.1016/0001-8708(81)90056-6}
  {\path{doi:http://dx.doi.org/10.1016/0001-8708(81)90056-6}}.

\bibitem{GNC2}
A.~Connes, An analogue of the thom isomorphism for crossed products of a c∗
  algebra by an action of r, Advances in Mathematics 39~(1) (1981) 31 -- 55.
\newblock \href
  {http://dx.doi.org/http://dx.doi.org/10.1016/0001-8708(81)90056-6}
  {\path{doi:http://dx.doi.org/10.1016/0001-8708(81)90056-6}}.

\bibitem{GNC3}
B.~Malekolkalami, M.~Farhoudi, Noncommutativity effects in \{FRW\} scalar field
  cosmology, Physics Letters B 678~(2) (2009) 174 -- 180.
\newblock \href
  {http://dx.doi.org/http://dx.doi.org/10.1016/j.physletb.2009.06.023}
  {\path{doi:http://dx.doi.org/10.1016/j.physletb.2009.06.023}}.

\bibitem{MT}
T.~Banks, W.~Fischler, S.~H. Shenker, L.~Susskind, $m$ theory as a matrix
  model: A conjecture, Phys. Rev. D 55~(0) (1997) 5112--5128.
\newblock \href {http://dx.doi.org/10.1103/PhysRevD.55.5112}
  {\path{doi:10.1103/PhysRevD.55.5112}}.

\bibitem{MT1}
E.~Witten, String theory dynamics in various dimensions, Nuclear Physics B
  443~(1–2) (1995) 85 -- 126.
\newblock \href
  {http://dx.doi.org/http://dx.doi.org/10.1016/0550-3213(95)00158-O}
  {\path{doi:http://dx.doi.org/10.1016/0550-3213(95)00158-O}}.

\bibitem{MSS}
G.~D. Barbosa, Noncommutative conformally coupled scalar field cosmology and
  its commutative counterpart, Phys. Rev. D 71~(14) (2005) 063511.
\newblock \href
  {http://dx.doi.org/http://dx.doi.org/10.1103/PhysRevD.71.063511}
  {\path{doi:http://dx.doi.org/10.1103/PhysRevD.71.063511}}.

\bibitem{MSS1}
H.~Garc\'{i}a-Compe\'an, O.~Obreg\'on, C.~Ram\'{i}rez, Noncommutative quantum
  cosmology, Phys. Rev. Lett. 88~(4) (2002) 161301.
\newblock \href
  {http://dx.doi.org/http://dx.doi.org/10.1103/PhysRevLett.88.161301}
  {\path{doi:http://dx.doi.org/10.1103/PhysRevLett.88.161301}}.

\bibitem{PHS}
C.~Bastos, O.~Bertolami, N.~C. Dias, J.~a.~N. Prata, Phase-space noncommutative
  quantum cosmology, Phys. Rev. D 78~(10) (2008) 023516.
\newblock \href
  {http://dx.doi.org/http://dx.doi.org/10.1103/PhysRevD.78.023516}
  {\path{doi:http://dx.doi.org/10.1103/PhysRevD.78.023516}}.

\bibitem{MALEK}
B.~Malekolkalami, K.~Atazadeh, B.~Vakili, Late time acceleration in a
  non-commutative model of modified cosmology, Physics Letters B 739~(0) (2014)
  400 -- 404.
\newblock \href
  {http://dx.doi.org/http://dx.doi.org/10.1016/j.physletb.2014.11.003}
  {\path{doi:http://dx.doi.org/10.1016/j.physletb.2014.11.003}}.

\bibitem{MP}
N.~Seiberg, E.~Witten, String theory and noncommutative geometry, Journal of
  High Energy Physics 1999~(09) (1999) 032.
\newblock \href {http://dx.doi.org/10.1088/1126-6708/1999/09/032}
  {\path{doi:10.1088/1126-6708/1999/09/032}}.

\bibitem{RM}
R.~Myrzakulov, \href{http://www.mdpi.com/1099-4300/14/9/1627}{Cosmology of f(t)
  gravity and k-essence}, Entropy 14~(9) (2012) 1627.
\newblock \href {http://dx.doi.org/10.3390/e14091627}
  {\path{doi:10.3390/e14091627}}.
\newline\urlprefix\url{http://www.mdpi.com/1099-4300/14/9/1627}

\bibitem{R2}
G.~Kofinas, E.~Papantonopoulos, E.~N. Saridakis, Self-gravitating spherically
  symmetric solutions in scalar-torsion theories, Phys. Rev. D 91~(14) (2015)
  104034.
\newblock \href
  {http://dx.doi.org/http://dx.doi.org/10.1103/PhysRevD.91.104034}
  {\path{doi:http://dx.doi.org/10.1103/PhysRevD.91.104034}}.

\bibitem{R3}
C.-Q. Geng, C.-C. Lee, E.~N. Saridakis, Observational constraints on
  teleparallel dark energy, Journal of Cosmology and Astroparticle Physics
  2012~(01) (2012) 002.
\newblock \href {http://dx.doi.org/10.1088/1475-7516/2012/01/002}
  {\path{doi:10.1088/1475-7516/2012/01/002}}.

\bibitem{17M}
T.~Curtright, D.~Fairlie, C.~Zachos, Features of time-independent wigner
  functions, Phys. Rev. D 58~(14) (1998) 025002.
\newblock \href
  {http://dx.doi.org/http://dx.doi.org/10.1103/PhysRevD.58.025002}
  {\path{doi:http://dx.doi.org/10.1103/PhysRevD.58.025002}}.

\bibitem{WD}
J.~Feinberg, Y.~Peleg, Self-adjoint wheeler-dewitt operators, the problem of
  time, and the wave function of the universe, Phys. Rev. D 52~(0) (1995)
  1988--2000.
\newblock \href {http://dx.doi.org/http://dx.doi.org/10.1103/PhysRevD.52.1988}
  {\path{doi:http://dx.doi.org/10.1103/PhysRevD.52.1988}}.

\bibitem{9M}
L.~Pimentel, C.~Mora, Noncommutative quantum cosmology, General Relativity and
  Gravitation 37~(4) (2005) 817--821.
\newblock \href {http://dx.doi.org/http://dx.doi.org/10.1007/s10714-005-0066-3}
  {\path{doi:http://dx.doi.org/10.1007/s10714-005-0066-3}}.

\end{thebibliography}

\end{document}